\documentclass[reprint,amsmath,amssymb,amsfonts,aps,prb,superscriptaddress,longbibliography]{revtex4-1}
\usepackage{graphicx}
\usepackage{bm}
\usepackage{epsfig}
\usepackage{epstopdf}
\usepackage{xcolor}
\usepackage{hyperref}
\hypersetup{colorlinks=true,allcolors=blue}
\usepackage{natbib}

\usepackage[normalem]{ulem}

\graphicspath{{figures/}}

\begin{document}
\title{Magnetic field induced quantum phases in a tensor network study of Kitaev magnets}

\author{Hyun-Yong Lee}
\affiliation{Institute for Solid State Physics, University of Tokyo, Kashiwa, Chiba 277-8581, Japan}
\affiliation{Department of Display and Semiconductor Physics, Korea University, Sejong 339-700, Republic of Korea}

\author{Ryui Kaneko}
\affiliation{Institute for Solid State Physics, University of Tokyo, Kashiwa, Chiba 277-8581, Japan}

\author{Li Ern Chern}
\affiliation{Department of Physics, University of Toronto, Toronto, Ontario M5S 1A7, Canada}

\author{Tsuyoshi Okubo}
\affiliation{Department of Physics, University of Tokyo, Tokyo 113-0033, Japan}

\author{Youhei Yamaji}
\affiliation{Department of Applied Physics, University of Tokyo,  Tokyo 113-8656, Japan}

\author{Naoki Kawashima}
\affiliation{Institute for Solid State Physics, University of Tokyo, Kashiwa, Chiba 277-8581, Japan}

\author{Yong Baek Kim}
\email{ybkim@physics.utoronto.ca}
\affiliation{Department of Physics, University of Toronto, Toronto, Ontario M5S 1A7, Canada}
\affiliation{Perimeter Institute for Theoretical Physics, Waterloo, Ontario N2L 2Y5, Canada}

\date{\today}

\begin{abstract}
Recent discovery of the half quantized thermal Hall conductivity in $\alpha$-RuCl$_3$, a candidate material for the Kitaev spin liquid, suggests the presence of a highly entangled quantum state in external magnetic fields. This field induced phase appears between the low field zig-zag magnetic order and the high field polarized state. Motivated by this experiment, we study possible field induced quantum phases in theoretical models of the Kitaev magnets, using the two dimensional tensor network approach or infinite tensor product states. We find various quantum ground states in addition to the chiral Kitaev spin liquid occupying a small area in the phase diagram. They form a band of emergent quantum phases in an intermediate window of external magnetic fields, somewhat reminiscent of the experiment. We discuss the implications of these results in view of the experiment and previous theoretical studies.
\end{abstract}
\maketitle

\noindent \textbf{\large Introduction} \\
Finding an unambiguous experimental evidence for quantum spin liquid has been a great challenge in the study of topological phases of matter\cite{Lucile2017, Zhou17}. Spin excitation spectra in quantum spin liquids, for example, consist of multiple excitations of underlying quasiparticles, namely spinons. Hence such spectra form a continuum and have no sharp excitations, which poses an inherent difficulty in identifying quantum spin liquids. In this context, the recent observation of half-quantized thermal Hall conductivity in the material $\alpha$-RuCl$_3$ in an external magnetic field is a remarkable discovery\cite{Kasahara18}. $\alpha$-RuCl$_3$ is a promising candidate for the gapless Kitaev spin liquid\,(KSL)\cite{Khaliullin2005, Jackeli2009, Plumb2014, Sears15, Johnson15, Kim15, Kim16, Yadav16, Zhou17, Banerjee2016, Luke16, Sinn16, Winter16, Leahy17a, Trebst2017, Banerjee17, Catuneanu18, Gohlke18, Winter18, Banerjee2018, Balz19, Wang19}, which is the ground state of an exactly solvable spin model\cite{Kitaev2006}. In the presence of magnetic field, it becomes the gapped chiral Kitaev spin liquid, which supports the chiral Majorana edge mode\cite{Kitaev2006}. The half-quantized thermal Hall
conductivity can be regarded as a unique signature of this Majorana edge state.

Without magnetic field, however, $\alpha$-RuCl$_3$ develops the zig-zag magnetic order at low temperatures\cite{Sears15, Johnson15, Kim15, Kim16, Yadav16}. Clearly, this must be due to the presence of spin 
interactions beyond the exactly solvable Kitaev model. A number of theoretical models are proposed and some minimal choices are the $K$-$\Gamma$-$\Gamma'$ and $K$-$\Gamma$-$J_3$ models. Here $K$ is the ferromagnetic Kitaev interaction, and $\Gamma$ is the bond dependent anisotropic interaction\cite{Kim16}. It is shown that a substantial $\Gamma$ is necessary to explain the large anisotropy of the magnetic susceptibility seen in experiments. The zig-zag order (ZZ) arises due to another anisotropic interaction $\Gamma'$, which is induced by the trigonal distortion of Cl octahedra, or the third neighbor antiferromagnetic Heisenberg interaction $J_3$\cite{Wang17}.
Hence the central question is how the zig-zag order would give
away to the chiral Kitaev spin liquid in the presence of magnetic field, and whether this happens in these minimal models.

\begin{figure}[!b]
	\includegraphics[width=0.49\textwidth]{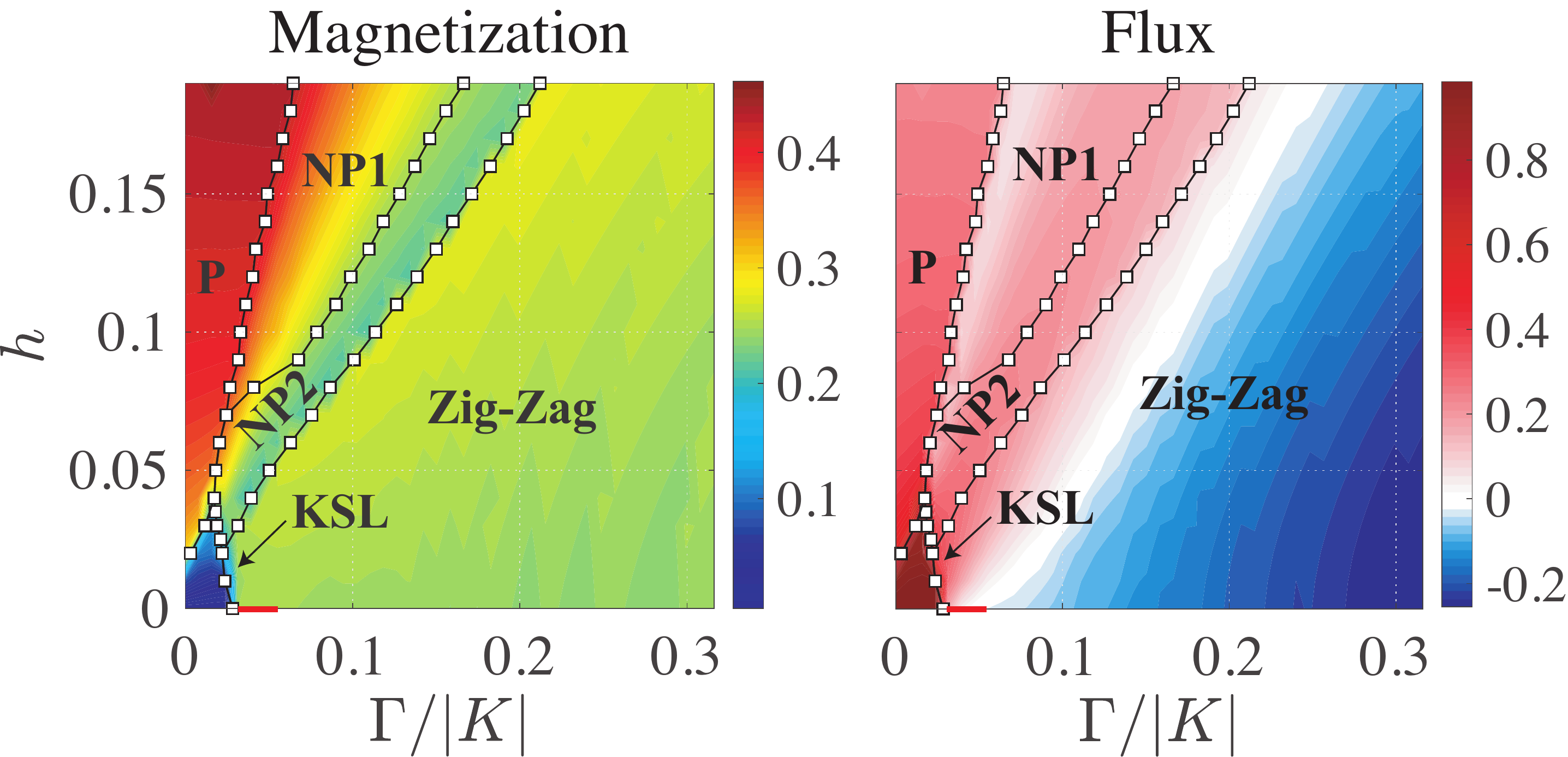}
	\caption{Ground state phase diagram. (a) The magnetization and (b) the flux expectation value of the $K$-$\Gamma$-$\Gamma'$ model. Here, KSL stands for the chiral Kitaev spin liquid, P for a spin-polarized, and NPs for nematic paramagnet phases, respectively. The red solid line at $h=0$ denotes a ferromagnetic phase\,(see text for details).}
	\label{fig:d6_phase}
\end{figure}

Previous results on the $K$-$\Gamma$-$\Gamma'$ model\cite{Kee19} obtained from exact diagonalization\,(ED) on the 24-site cluster (when the magnetic field is tilted away from $[111]$ direction so that $C_3$ rotation symmetry is explicitly broken), and density matrix renormalization group\,(DMRG) on the 2-leg ladder geometry suggest that the chiral Kitaev spin liquid is stabilized in a large window of magnetic field and $\Gamma/K$ between the zig-zag and polarized phases. Another recent theoretical work on the classical model\cite{Li19}, however, shows that there exist a multitude of complex magnetic orders with large unit cells in a similar window of intermediate magnetic fields. Many of these phases cannot be accommodated in small systems used in the ED and DMRG calculations mentioned above. In order to resolve this issue, theoretical studies of the quantum model in the thermodynamic limit are necessary.

In this article, we present the results of the infinite Tensor Product States (iTPS) studies on the $K$-$\Gamma$-$\Gamma'$ and $K$-$\Gamma$ models, which directly deal with the two dimensional thermodynamic limit. Here we can treat the Kitaev spin liquid and the classical complex magnetically ordered states on equal footing. Our study shows that the Kitaev spin liquid only occupies a small corner in the magnetic field phase diagram. On the other hand, we find novel quantum phases, namely the nematic paramagnets, to be emergent in an intermediate window of magnetic fields. Apart from providing a definite prediction for $\alpha$-RuCl$_3$, our result also addresses the general question as to what are the possible quantum phases around the Kitaev spin liquid in other spin-orbital entangled honeycomb magnets, which are described by similar theoretical models containing substantial $K$ and $\Gamma$ interactions. Below, we explain the phase diagram and discuss the nature of magnetic field induced quantum phases. \\

\noindent \textbf{\large Results} \\
\noindent \textbf{Model.} We begin with the Hamiltonian of the $K$-$\Gamma$-$\Gamma'$ model\cite{Kee14, Kee19, Li19}: $\hat{H} = \sum_{\langle ij \rangle_{\gamma}} \hat{H}_{ij}^\gamma$ with 
\begin{align}
	\hat{H}_{ij}^\gamma & =  - \frac{\bf h}{3} \cdot ({\bf S}_i + {\bf S}_j) + K S_i^\gamma S_j^\gamma + \Gamma \left( S_i^\mu S_j^\nu + S_i^\nu S_j^\mu \right) \nonumber\\
	& + \Gamma'\left( S_i^\mu S_j^\gamma + S_i^\gamma S_j^\mu + S_i^\nu S_j^\gamma + S_i^\nu S_j^\gamma \right),
	\label{eq:hamiltonian}
\end{align}
where $\langle ij\rangle_\gamma$ denotes the pair of the nearest neighbor sites, $i$ and $j$, on the $\gamma$-bond with $\gamma = x,y,z$ as depicted in the Methods section. The $K$ term is the isotropic Kitaev interaction. Here, $(\gamma, \mu, \nu)$ forms a cyclic permutation of $(x,y,z)$ such that off-diagonal spin exchanges are represented by $\Gamma$ and $\Gamma'$ interactions. In both classical\cite{Li19} and quantum\cite{Kee19} limits, a small $\Gamma'$ interaction induces the ZZ magnetic order at small magnetic fields, which gives away to other competing phases at larger magnetic fields. Throughout this article, we fix $\Gamma'=-0.03$ in units of $\sqrt{K^2 +\Gamma^2}=1$ and focus on the ferromagnetic Kitaev and antiferromagnetic $\Gamma$ interactions, i.e., $K<0$ and $\Gamma>0$, which is relevant to the material $\alpha$-RuCl$_3$. The magnetic field is applied along the $[111]$ direction, i.e., ${\bf h} = h(1,1,1)/\sqrt{3}$. We also consider the effect of tilting the magnetic field from the $[111]$ direction.
The Hamiltonian is invariant under the transformation $C_6 U_{C_6}$: $[C_6 U_{C_6}, H]=0$ where $C_6$ denotes $60^\circ$ lattice rotation about the center of the plaquette while $U_{C_6}$ cyclically permutes the components of the spin operator, i.e., $S^x \rightarrow S^y \rightarrow S^z \rightarrow S^x$. For simplicity, we refer to this symmetry as the {\it rotational} symmetry. \\

\begin{figure}[!t]
	\includegraphics[width=0.49\textwidth]{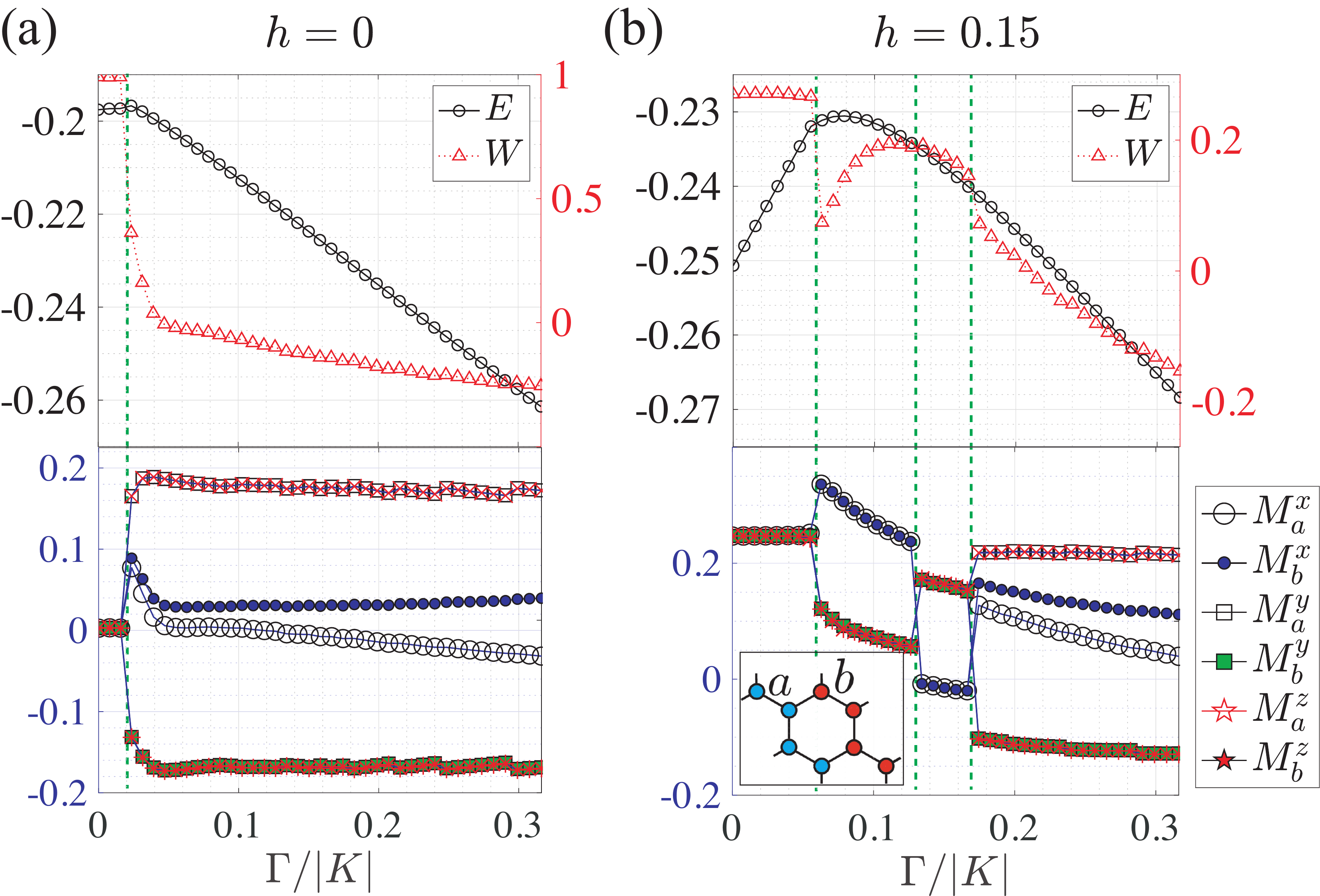}
	\caption{Distinguishing phases. Plots of variational energy $E$, flux $W$\,(upper) and components of magnetization $M_{a,b}^\gamma$\,(lower) at (a) $h=0$, (b) $h=0.15$ as a function of $\Gamma/|K|$ with $\Gamma'=-0.03$. The green dotted lines specify the phase boundaries. The information on numerical parameters is given in the Methods section.}
	\label{fig:magnetization}
\end{figure}

\noindent \textbf{Identification of each phase.} To determine the phase boundaries and characterize each phase, we optimize the iTPS using the imaginary time evolution\,(ITE)\cite{Tao08} and measure the energy density $E =  \langle H\rangle / N_s$, magnetization $M \equiv N_s^{-1} \sum_i^{N_s} \sqrt{ \langle {\bf S}_i \rangle^2 }$ and flux $W \equiv N_p^{-1} \sum_p^{N_p} \langle \hat{W}_p \rangle$. Here, $\hat{W}_p = \hat{\sigma}_1^x \hat{\sigma}_2^y \hat{\sigma}_3^z \hat{\sigma}_4^x \hat{\sigma}_5^y \hat{\sigma}_6^z$ is the flux operator\cite{Kitaev2006} on a plaquette $p$, the site indices $1\!\!-\!\!6$ are defined in the Methods section and $N_{s(p)}$ is the number of sites\,(plaquettes) in the system. As shown in the phase diagram Fig.\,\ref{fig:d6_phase}, we identify five distinct phases, i.e., KSL, polarized\,(P), nematic paramagnetic\,(NP1 and NP2) and ZZ phases in the parameter region $0<\Gamma/|K| \leq 0.3$ and $0<h \leq 0.2$. \\

\noindent \textbf{Small extent of Kitaev spin liquid in field.} First, the KSL ground state survives only in a small corner of the phase diagram. In the KSL phase, the magnetization and the fluctuation of vortices are suppressed, i.e., $M \ll 1/2$ and $W \approx 1$ as shown in Figs.\,\ref{fig:d6_phase} and \ref{fig:magnetization}\,(a). It disagrees with the largely extended KSL phase observed in the 24-site ED and DMRG studies on the two-leg ladder system in Ref.\,\onlinecite{Kee19}. The discrepancy may imply that taking the thermodynamic limit is important. At zero field, there is a transition from KSL to a ferromagnetic\,(FM) phase where spins are aligned in the $[1\bar{1}\bar{1}]$ direction\,(red solid line in Fig.\,\ref{fig:d6_phase}). However, with a very weak magnetic field\,($h=0.005$), the FM phase disappears, and a direct phase transition from KSL to ZZ occurs. With increasing $h$, the transition from KSL to the P phase occurs at a finite $h$, where spins start aligning in the $[111]$ direction. The fate of the FM states will be discussed in details later. We have found that the field induced phase transition with $(\Gamma,\Gamma') = (0,-0.03)$ occurs at $h_c^{K\Gamma'} \approx 0.011$, which is smaller than $h_c^{K} \approx 0.02$\cite{Jiang11, Zhu18, Gohlke18a, Hickey19, Kaib19} of the pure Kitaev model\,(see Supplementary Note 5). \\

\begin{figure}[!t]
	\includegraphics[width=0.49\textwidth]{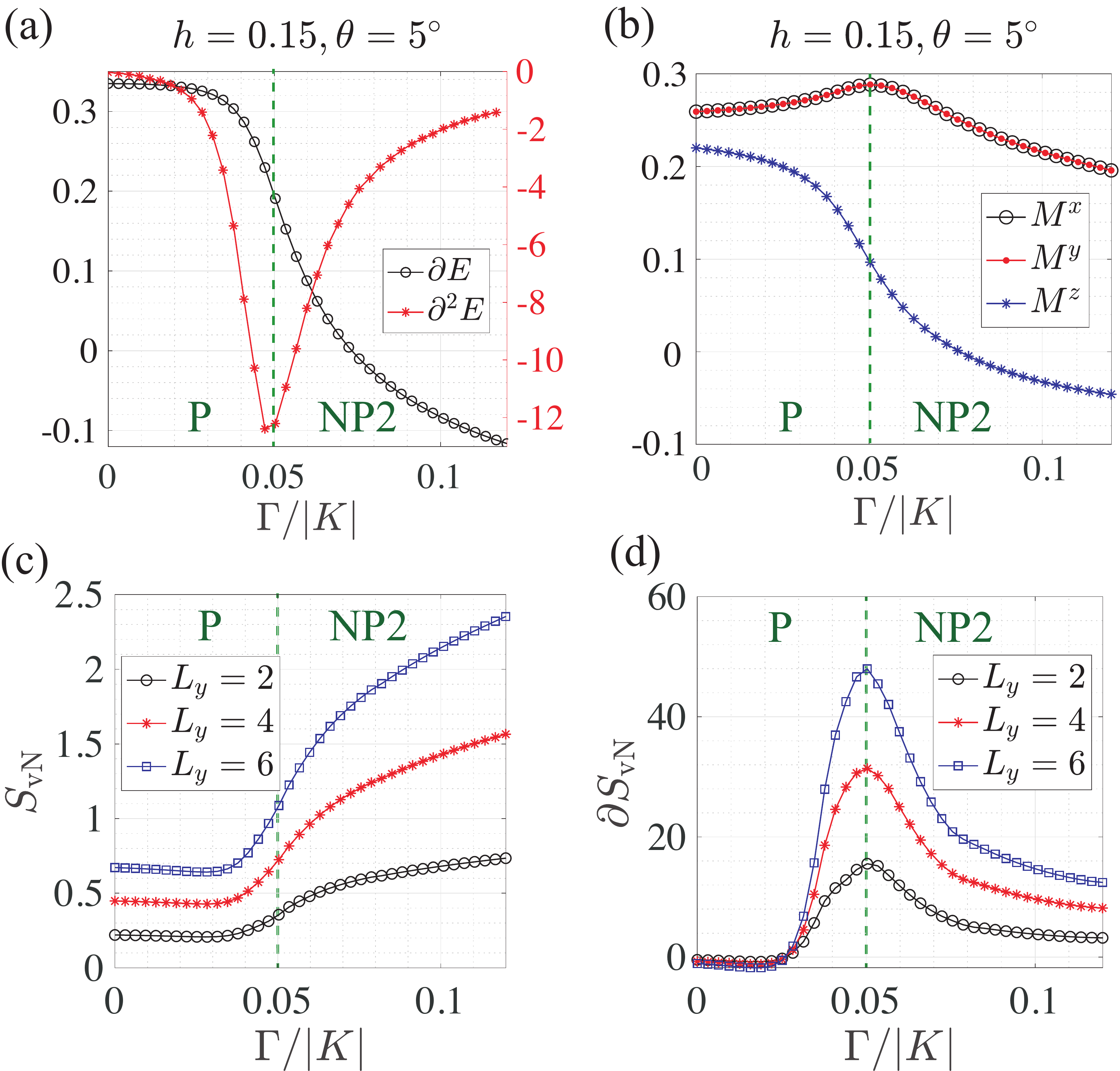}
	\caption{Transition between the P and NP2 phases. Plots of (a) the first and second derivatives of the energy density, (b) magnetization components, (c) entanglement entropy\,($S_{\rm vN}$) and (d) its first derivative with the tilting angle $\theta = 5^\circ$ and $h=0.15$. Here, $S_{\rm vN}$ is measured on the cylinder geometry with the circumference $L_y$. The information on numerical parameters is given in the Methods section.}
	\label{fig:diff_e}
\end{figure}

\noindent \textbf{Nematic paramagnetic phases.} As $\Gamma$ increases, the magnetic field induced phase is no longer the Kitaev spin liquid. The ZZ order gives away to interesting intermediate phases NP1 and NP2 \,(Fig.\,\ref{fig:d6_phase}) before the system enters the P phase at high field. Both phases are nematic in the sense that the rotational symmetry is spontaneously broken down to the $C_2$ rotational symmetry. More specifically, the local energy $E^\gamma=\langle \hat{H}_{ij}^\gamma\rangle $ depends on the direction of bond $\gamma$: $E^x < E^y = E^z$ in NP1 while $E^x > E^y = E^z$ in NP2. It also leads to the anisotropic magnetization, i.e., $M^x \neq M^y = M^z$ {\it etc}, as presented in Fig.\,\ref{fig:magnetization}\,(b). 

In the classical limit, the 8-site, 18-site, and 32-site magnetic orders are stabilized in a similar parameter regime\cite{Li19}. Our result indicates that strong quantum fluctuation melts the competing large unit-cell orders, leading to the restoration of the translational symmetry, while the rotational symmetry remains broken. We have also found that the NP phases appear and survive down to almost zero field limit in the $K$-$\Gamma$ model as shown below. By increasing the accuracy of the iTPS representation\,(see Supplementary Note 7), we have confirmed that the NP states are quantum paramagnet and develop finite magnetization only in the presence of the field. 

In the [111] magnetic field, the nature of the transition between P and NP1 phases is not clear. Even though the local observables show finite jumps at the transition, these are not very distinctive compared to other transitions and may originate from the inherently biased optimization in ITE, which is analyzed carefully in Supplementary Note 2. The non-triviality of the NP phases is revealed by tilting the magnetic field slightly towards the $[11\bar{2}]$ direction. Fig.\,\ref{fig:diff_e}\,(a) presents the optimized energy and its second derivative with respect to $\Gamma/|K|$ at the tilting angle $\theta = 5^\circ$. Notice that, due to the tilted field, the model breaks the rotational symmetry explicitly, and thus there is no remaining symmetry discriminating the P and NP phases. Nevertheless, the second derivative of the energy strongly suggests a continuous phase transition between the P and NP2 phases\,[see Fig.\,\ref{fig:diff_e}\,(b)] at $\Gamma/|K|\approx 0.05$. Note that the tilted field with $\theta>0$ leads to a transition from the P phase directly to the NP2 phase. On the other hand, tilting the field in the opposite direction\,($\theta < 0$) favors the NP1 phase and therefore gives rise to a transition from the P phase to the NP1 phase\,(see Supplementary Note 6). The continuous nature of these transitions can be seen even more clearly in the entanglement entropy\,(EE)\cite{Li16, Liu16, Cho17}. The boundary theory of TPS\cite{Cirac11} has been employed to measure the EE on the cylinder geometry with the circumference $L_y$, and the result is presented in Fig.\,\ref{fig:diff_e}\,(c) and (d). The NP1 state is highly entangled and its EE increases with $\Gamma$, while the P state has a low and constant EE. The first derivative of the EE exhibits a peak at the same point as that of the second derivative of the energy, and it becomes sharper with increasing $L_y$ and the accuracy of the variational ansatz\,(see Supplementary Note 3). Therefore, we conclude that there is a continuous transition between the P and NP2 phases at $\Gamma/|K| \approx 0.05$. As mentioned above, the P and NP phases cannot be distinguished by conventional symmetries, thus the continuous transition implies a topological phase transition from the trivial phase\,(P) to a topological or non-trivial phase\,(NP2). It is worth noting that the tilted field makes the numerical optimization much more stable as analyzed in Supplementary Figure 2. \\

\begin{figure}[!t]
  \includegraphics[width=0.49\textwidth]{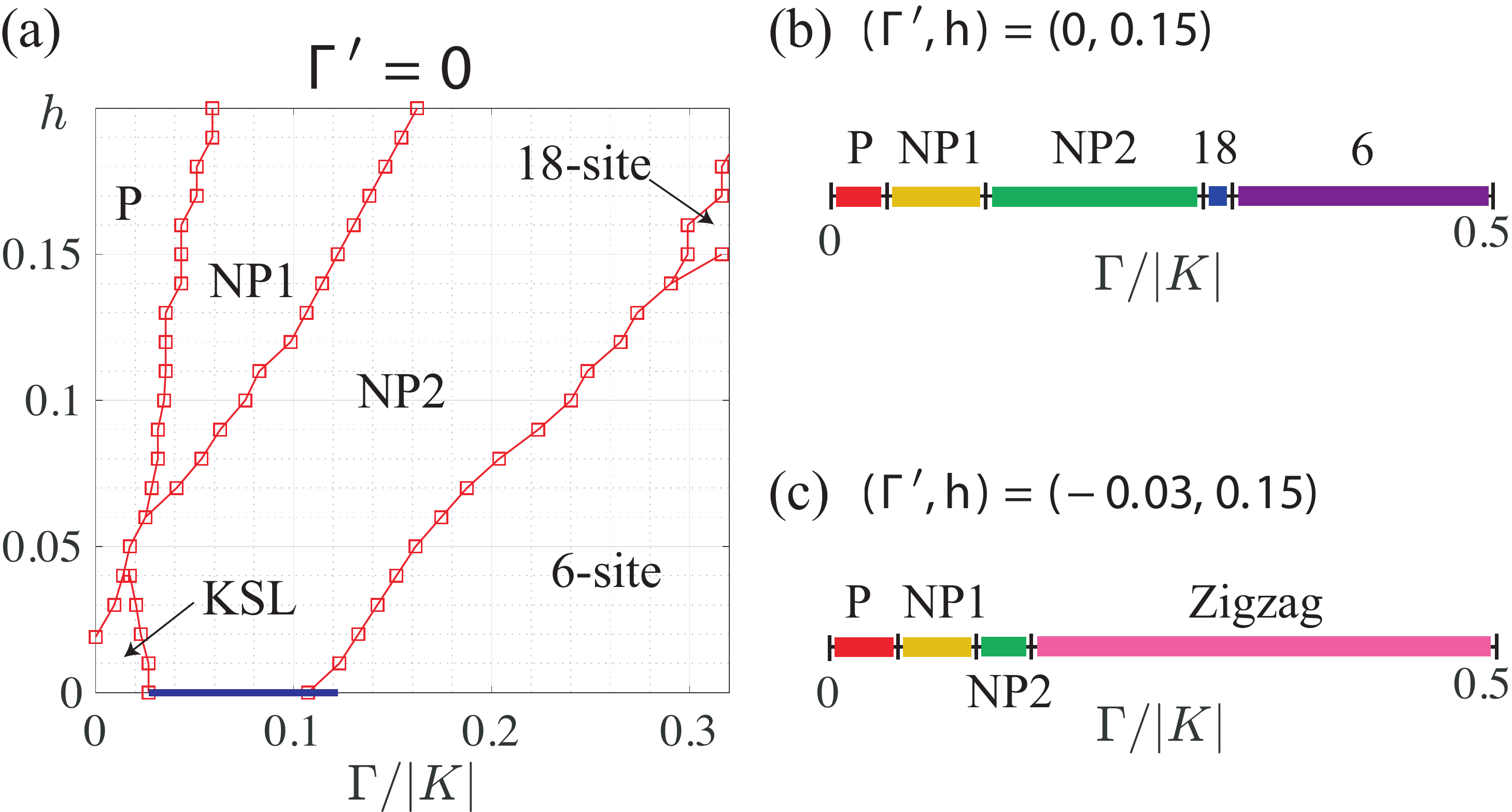}
  \caption{Comparison to the $K-\Gamma$ model. (a) Phase diagram at $\Gamma'=0$ as functions of $\Gamma/|K|$ and $h$. NP1 and NP2 phases are larger than those at $\Gamma'=-0.03$ since the zig-zag phase is less dominant. Complexed magnetic phases appear when $\Gamma/|K| \gtrsim 0.3$ and $h\gtrsim 0.15$. The blue solid line at $h=0$ denotes a ferromagnetic phase\,(see text for details). Larger $\Gamma$ phase diagrams at (b) $(\Gamma',h) = (0,0.15)$ and (c) $(\Gamma',h) = (-0.03,0.15)$. The information on numerical parameters is given in the Methods section.}
  \label{fig:Gp0_phase_diag}
\end{figure}

\noindent \textbf{$K$-$\Gamma$ model.} Finite $\Gamma'$ is responsible in stabilizing the ZZ order at low fields. When $\Gamma' = 0$, we may expect a more significant competition between various phases including the complex classical magnetic orders. We find that the NP phases are already present in the $K$-$\Gamma$ model as shown in the phase diagram in Fig.\,\ref{fig:Gp0_phase_diag}. On the other hand, the complex magnetic orders with large magnetic unit cells appear for sufficiently large $\Gamma$ (typically $\Gamma/|K|\gtrsim 0.3$). For example, the 6-site order phase appears at lower field $h\lesssim 0.15$
while the 18-site order phase appears at higher field $h\gtrsim 0.15$ as presented in Figs.\,\ref{fig:Gp0_phase_diag}\,(a) and (b). These are the same magnetic orders reported in the classical phase diagram\cite{Li19}. Quantum fluctuations seem to favor NP1 and NP2 phases at small $\Gamma$, and push the classical orders to the parameter region with larger $\Gamma$. As in the case of $\Gamma'=-0.03$, the FM phase appears between the KSL and the 6-site order at $h=0$. This reminds us of a ferromagnetic phase in a tiny area of the phase diagram in the variational Monte Carlo study in Ref.\,\onlinecite{Wang19}. However, the NP2 state is almost degenerate with FM phase in this region, i.e., the energy difference is only $\Delta E \sim O(10^{-4})$. Moreover, even this tiny energy difference is decreasing as the accuracy of the iTPS further increases\,(see Supplementary Note 8). With these results and given that the FM quickly loses to NP2 with a very small $h$, NP2 may become a stable ground state or degenerate with the FM phase at $h=0$ as the variational iTPS approaches to the exact ground state. The NP phases are reminiscent of the $K\Gamma$ spin liquid reported in the previous iDMRG study\cite{Gohlke18}, where the rotational symmetry is broken in similar manner. \\



\noindent \textbf{\large Conclusion} \\
We have used iTPS optimization to investigate the field induced quantum phases in the $K$-$\Gamma$-$\Gamma'$ model. 
Apart from the well established chiral KSL, we discover the stabilization of the nematic paramagnets NP1 and NP2 at intermediate magnetic fields. The NP phases break lattice rotational symmetry spontaneously, and take place between the low field ZZ magnetic order and the high field polarized state. In contrast to the previous 24-site ED and 2-leg ladder DMRG study\cite{Kee19}, the KSL is found to survive only in a small corner of the phase diagram. Instead, the NP phases occupy a large portion of the phase diagram and hence are more likely to be observed. We propose that, to probe the nematic paramagnets experimentally, one could measure longitudinal thermal conductivity and magnetic susceptibility over the in-plane directions, which would display the breaking of $C_3$ symmetry. We also find that the NP phases are already present in the $K$-$\Gamma$ model in zero and finite magnetic field. The NP phases in the $K$-$\Gamma$ model give away to the complex magnetic orders with large unit cells when $\Gamma/|K|$ becomes large, making contact with the classical phase diagram reported earlier.

In order to clarify the nature of the NP phases, we examine the effect of tilting the magnetic field\,($\theta=5^\circ$ from the [111] direction).
Here the transition between the polarized (P) and NP2 phases is continuous,
judging from the singular behaviors in the second derivative of the energy and the first derivative of the entanglement entropy. Since $C_3$ is broken in both of the P and NP2 phases in the tilted field, the continuous transition would imply that NP2 is not a trivial product state. This leaves the interesting possibility that the NP phases are non-trivial topological states. The precise nature and thermal Hall response of these states would be interesting subjects of future study. \\

\begin{figure}[!t]
	\includegraphics[width=0.49\textwidth]{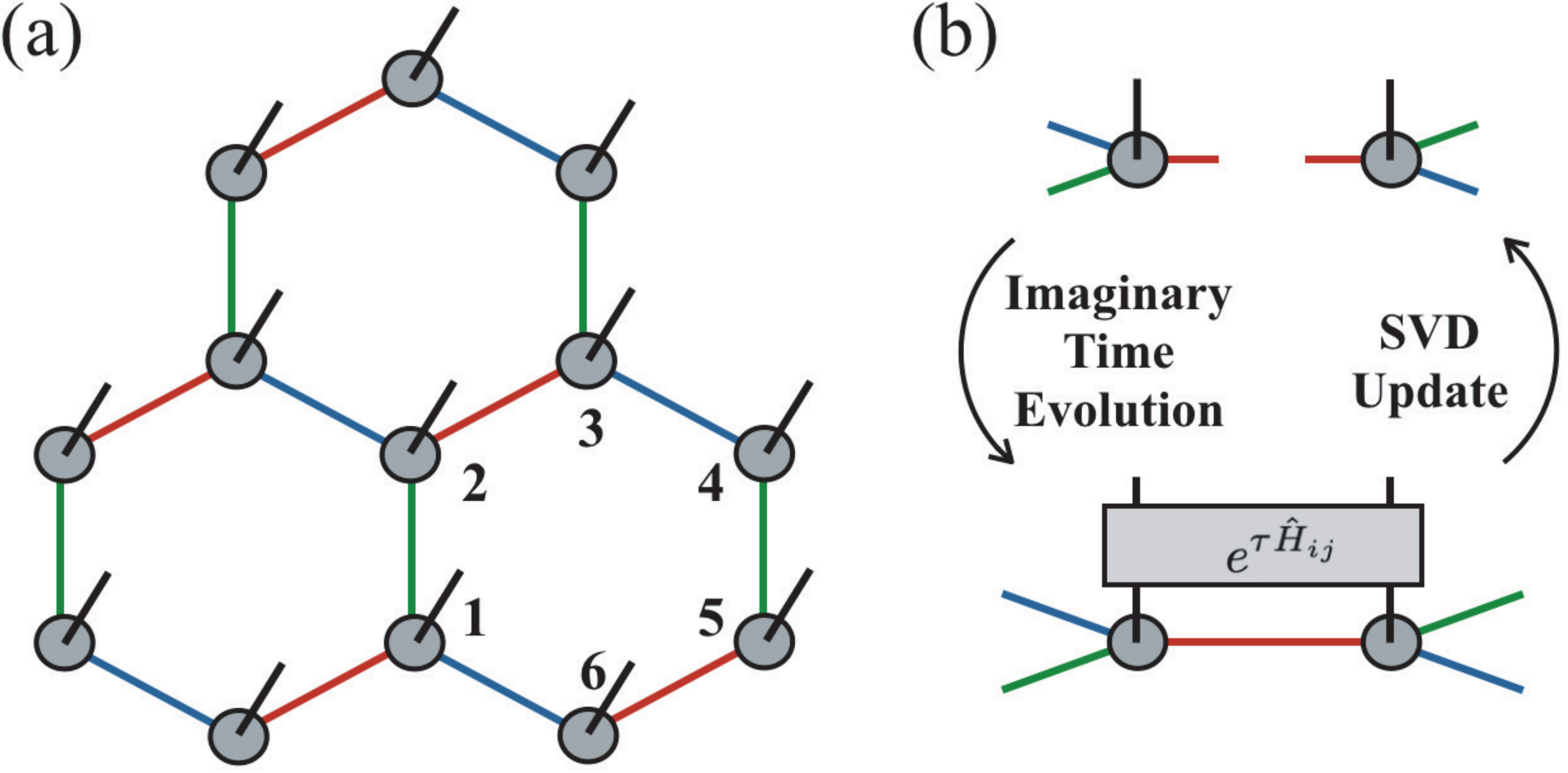}
	\caption{Methods. Schematic figures of (a) the honeycomb TPS where the $x$-, $y$- and $z$-bonds defined in Eq.\,\eqref{eq:hamiltonian} are specified by red, blue and green colors, respectively, and (b) the optimization and update processes of the local tensor, where the black solid line denotes the physical degrees of freedom and $e^{-\tau \hat{H}_{ij}}$ is the local imaginary time evolution operator\,(see text for detail).}
	\label{fig:schematic}
\end{figure}

\noindent \textbf{\large Methods} \\
\noindent \textbf{Infinite tensor product states and imaginary time evolution.} In order to carve out the ground state phase diagram, we employ the iTPS representation\cite{Verstraete08} on the honeycomb lattice and optimize it with respect to the Hamiltonian in Eq.\,\eqref{eq:hamiltonian}. The iTPS wavefunction $\psi_{\{ s_i \}} = {\rm tTr} \prod_i [T_i]_{\alpha_i \beta_i \gamma_i }^{s_i} $ is illustrated in Fig.\,\ref{fig:schematic}\,(a), where $s_i$ denotes the spin state at site $i$, and ${\rm tTr}$ represents the trace over the virtual indices $(\alpha_i,\beta_i,\gamma_i)$ of the local tensor $T_i$. The accuracy of the iTPS representation becomes better as the dimension of the virtual indices, or the bond dimension $D$, increases. The ITE is adopted for optimization, i.e. the two-site gate $e^{-\tau \hat{H}_{ij}^\gamma }$ is applied on every bond with fixed $\tau=0.01$. Then, the local tensors are updated by the singular value decomposition\cite{Tao08}. Iterating this two-step procedure\,(Fig.\,\ref{fig:schematic}\,(b)) drives the initial state into the ground state.

Since the ITE with such a simple update can be easily biased by the initial choice of $T_i$, we optimize various trial states and choose the lowest energy state as the ground state. We consider the string gas\,(SG) representation of the KSL in Ref.\,\onlinecite{HY19} and the classical magnetic orders found in Ref.\,\onlinecite{Li19}. Note that the ITE starting from the SG state, i.e., $|\psi\rangle \approx \left( e^{-\tau \hat{H}} \right)^{\mathcal{N}}|\rm{SG}\rangle$, provides the lowest energy states near the pure Kitaev limit, which allows us to determine the KSL phase. We also include the ferromagnetic [111] state\,(FM[111]), where all spins are aligned in the $[111]$ direction, ZZ, 6-site and 18-site magnetic orders found in Ref.\,\onlinecite{Li19}. In addition, we use the FM[100], FM[011] and FM[1$\bar{1}\bar{1}$] states as other possible initial states. Details of the initial states are provided in Supplementary Note 1. Due to the complexity, we did not take into account the 32-site and 50-site magnetic order discovered in the classical phase diagram\cite{Li19}, which might be relevant for larger $\Gamma$ and $h$ than the parameter region considered in this work. To measure the physical quantities after the optimization, we employ the corner transfer matrix renormalization group\,(CTMRG) method\cite{Nishino96, Nishino98, Corboz2010}. The parallel C++ library mptensor\cite{mptensor} is utilized to perform CTMRG and ITE. The main results in this article are obtained with the bond dimension $D=6$. It turns out that the phase diagram and physical quantities do not change much as increasing $D$ as discussed in SI. \\

\noindent \textbf{\large Code availability} \\
All numerical codes in this paper are available from the authors upon reasonable request. \\

\noindent \textbf{\large Data availability}\\
All relevant data in this paper are available from the authors upon reasonable request. \\

\noindent \textbf{\large Acknowledgement} \\
We thank Matthias Gohlke and Hae-Young Kee for fruitful discussions. The computation was performed at the Supercomputer Center, ISSP, University of Tokyo, and also on K-computer (project-ID: hp190196). This work was supported by ImPACT Program of Council for Science, Technology and Innovation [Cabinet Office, Government of Japan] (N.K.) and MEXT as ``Exploratory Challenge on Post-K computer"\, [Frontiers of Basic Science: Challenging the Limits] (H.-Y.L.), and ``Priority Issue on Post-K computer" [Creation of New Functional Devices and High-Performance Materials to Support Next-Generation Industries] (R.K., T.O., and Y.Y.). T.O. was financially supported by JSPS KAKENHI 19K03740. Y.Y. was supported by PRESTO, JST (JPMJPR15NF). This work was further supported by the NSERC of Canada, and the Killam Research Fellowship of the Canada Council for the Arts (Y.B.K.). \\

\noindent \textbf{\large Author contributions} \\
H.-Y.L. and R.K. performed the major part of the tensor network computations. Y.B.K. conceived the project and supervised the study. H.-Y.L., R.K., L.E.C., T.O., Y.Y., N.K. and Y.B.K. participated in the design of the project/computations and contributed to the writing of the manuscript. \\

\noindent \textbf{\large Competing interests} \\
The authors declare no competing interests. \\

\noindent \textbf{\large Additional information} \\
\textbf{Supplementary information} is available for this paper. \\


%

\clearpage
\onecolumngrid
\begin{center}
\textbf{\large  Supplemental Material:}
\end{center}

\setcounter{equation}{0}
\setcounter{figure}{0}
\setcounter{table}{0}
\setcounter{page}{1}

\begin{center}
\parbox[t][4cm][s]{0.8\textwidth}{	We define the unit cell structures and the magnetic moment on each sublattice of the initial magnetic states used in the optimization. Then, the dependence of the optimized states on the initial state and the bond dimension are discussed. In addition, we provide details of the field induced phase transition, the critical field strength $h_c$ at $\Gamma=0$ and $-0.03$, and tilting the magnetic field towards the [001]-direction\,($\theta < 0$).}
\end{center}

\author{Hyun-Yong Lee}
\affiliation{Institute for Solid State Physics, University of Tokyo, Kashiwa, Chiba 277-8581, Japan}
\affiliation{Department of Display and Semiconductor Physics, Korea University, Sejong 339-700, Republic of Korea}

\author{Ryui Kaneko}
\affiliation{Institute for Solid State Physics, University of Tokyo, Kashiwa, Chiba 277-8581, Japan}

\author{Li Ern Chern}
\affiliation{Department of Physics, University of Toronto, Toronto, Ontario M5S 1A7, Canada}

\author{Tsuyoshi Okubo}
\affiliation{Department of Physics, University of Tokyo, Tokyo 113-0033, Japan}

\author{Youhei Yamaji}
\affiliation{Department of Applied Physics, University of Tokyo,  Tokyo 113-8656, Japan}

\author{Naoki Kawashima}
\affiliation{Institute for Solid State Physics, University of Tokyo, Kashiwa, Chiba 277-8581, Japan}

\author{Yong Baek Kim}
\affiliation{Department of Physics, University of Toronto, Toronto, Ontario M5S 1A7, Canada}
\affiliation{Perimeter Institute for Theoretical Physics, Waterloo, Ontario N2L 2Y5, Canada}
\date{\today}

\section*{Supplementary Note 1: Unit cell structures of initial magnetic states}

We present the details of the classical magnetic orders used as the initial states in the imaginary time evolution optimization. We have considered ten different initial states, which are the string gas, FM[111], FM[100], FM[011], zigzag, 6-site, 8-site, 18-site1, 18-site2 and 18-site3 states. The definition of the string gas state is given in Supplementary Ref.\,\onlinecite{HY19}. The FM$[abc]$ state denotes the classical product state where all spins are aligned in the $[abc]$-direction. The remaining states are classical product states defined on larger unit cells. The structure of the magnetic unit cells are depicted in Supplementary Fig.\,\ref{fig:unitcell}, and the (normalized) magnetization components are given below.

\begin{figure}[!b]
	\includegraphics[width=0.99\textwidth]{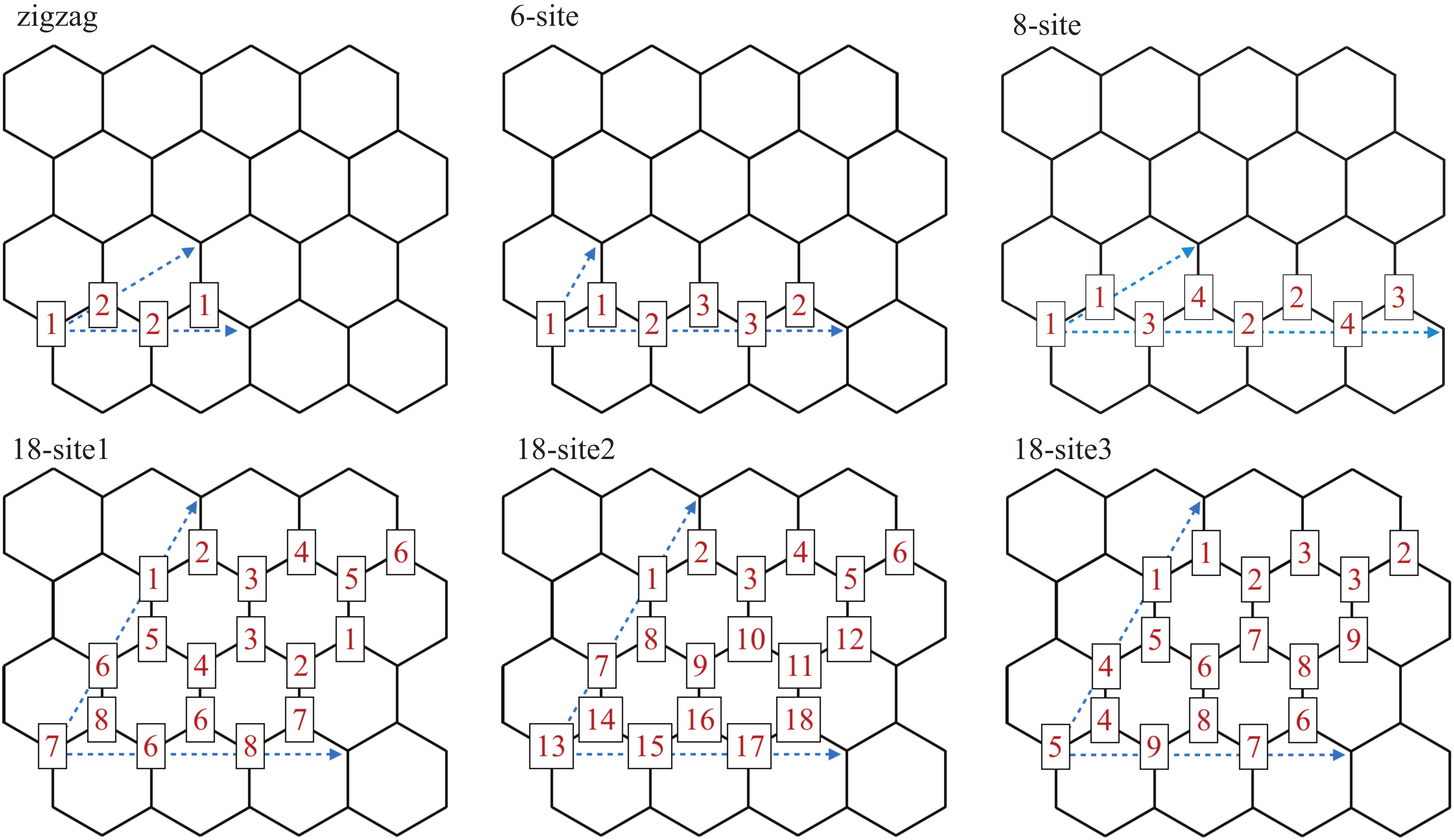}
	\caption{The unit cell structures of various magnetic states, where the sublattices are labeled with numbers, and the blue arrows denote the primitive vectors.}
	\label{fig:unitcell}
\end{figure}
\begin{itemize}
	\item zigzag\\ 1:$(0,1/\sqrt{2},1/\sqrt{2})$, \quad 2:$(0,-1/\sqrt{2},-1/\sqrt{2})$
	\item 6-site\\ 1:$(0,1/\sqrt{2},1/\sqrt{2})$, \quad 2:$(1/\sqrt{2},-1/2,-1/2)$ , \quad 2:$(-1/\sqrt{2},-1/2,-1/2)$ 
	\item 8-site\\ 1:$(1/\sqrt{2},1/\sqrt{2},0)$, \quad 2:$(-1/\sqrt{2},1/\sqrt{2},0)$, \quad 3:$(1/\sqrt{2},1/\sqrt{2},0)$, \quad 4:$(-1/\sqrt{2},-1/\sqrt{2},0)$
	\item 18-site1\\ 1: $(0.694426963314,  0.508808015180,  0.508808015180)$, \quad 2: $(0.917670696610,  0.280963069267,  0.280963069267)$, \quad 3: $(0.360919338886,  0.360919338886, -0.859927009482)$, \quad 4: $(0.280963069267,  0.917670696610,  0.280963069267)$, \quad 5: $(0.508808015180,  0.694426963314,  0.508808015180)$, \quad 6: $(0.508808015180,  0.508808015180,  0.694426963314)$, \quad 7: $(0.360919338886, -0.859927009482,  0.360919338886)$, \quad 8: $(0.280963069267,  0.280963069267,  0.917670696610)$
	\item 18-site2 \\1: $(0.358086657424,0.627886711621,0.691037063510)$, \quad2: $(0.317167993518,0.662906654162,0.678202942898)$, \quad 3: $(0.058719049073,0.953736826431,0.294852741526)$, \quad 4: $(0.076804072888,0.231146393012,0.969882714242)$, \quad 5: $(0.830779222896,0.194767830929,0.521412864090)$, \quad 6: $(0.691037063510,0.627886711621,0.358086657424)$, \quad 7: $(0.478273748206,0.828840613223,0.290305803678)$, \quad 8: $(0.521412864090,0.194767830929,0.830779222896)$, \quad 9: $(-0.969882714242,0.231146393012,0.0768040728)$, \quad 10: $(-0.917656299286,0.261691990911,0.2990388240)$, \quad 11: $(0.263587705748,0.179270501689,0.947831002132)$, \quad 12: $(0.290305803678,0.828840613223,0.478273748206)$, \quad 13: $(0.723791516161,0.349229309838,0.595117408823)$, \quad 14: $(0.947831002132,0.179270501689,0.263587705748)$, \quad 15: $(0.299038824024,0.261691990911,0.917656299286)$, \quad 16: $(0.294852741526,0.953736826431,0.058719049073)$, \quad 17: $(0.678202942898,0.662906654162,0.317167993518)$, \quad 18: $(0.595117408823,0.349229309838,0.723791516161)$
	\item 18-site3\\ 1: $(0.256377480005,0.705963966202,0.660216226831)$, \quad   2: $(0.877902927220,0.452899788357,0.155461352380)$, \quad   3: $(0.639181668028,0.427660602011,0.639181668028)$, \quad   4: $(0.660216226831,0.705963966202,0.256377480005)$, \quad   5: $(0.691497143275,0.208957894528,0.691497143275)$, \quad   6: $(0.693921708246,0.192211668868,0.693921708246)$, \quad   7: $(0.516130572975,0.834166479598,0.194359244588)$, \quad   8: $(0.194359244588,0.834166479598,0.516130572975)$, \quad   9: $(0.155461352380,0.452899788358,0.877902927220)$ 
\end{itemize}
\section*{ Supplementary Note 2: Initial state dependence of imaginary time evolution }
\begin{figure}[!h]
	\includegraphics[width=0.99\textwidth]{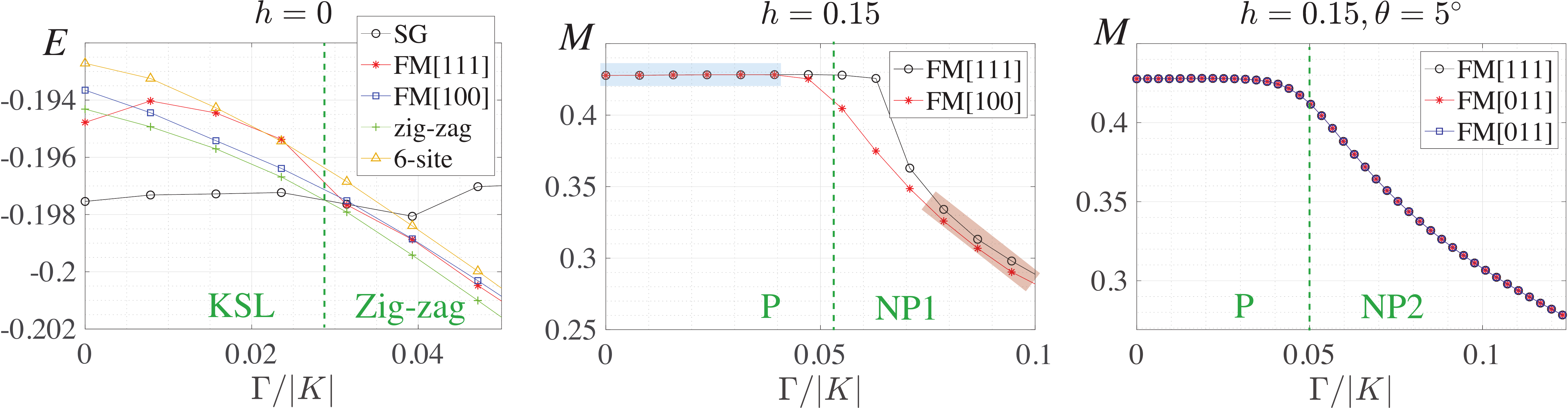}
	\caption{(left) The variational energies depending on various initial states at $(\Gamma',h)=(-0.03,0)$, the magnetizations of the optimized states obtained from (center) the initial FM[111], FM[100] states with $\theta = 0$ and (right) the initial FM[111], FM[011], FM[100] states with $\theta = 5^\circ$. The green dotted line indicates the phase boundary.}
	\label{fig:initial_dep}
\end{figure}

As mentioned in the main text, the imaginary time evolution method combined with the simple local update of tensors is easily biased by the choice of the initial state. Here, we discuss the initial state dependence of the optimized states. First, we find that the ground states in the chiral Kitaev spin liquid phase are obtained only from the string gas\,(SG) initial state\cite{HY19}, as shown in the left panel of Supplementary Fig.\,\ref{fig:initial_dep}. As one can see, the optimized states starting from the SG state has significantly lower energy than others, thus the choice of the initial state is very important in the KSL phase. Other magnetic initial states do not evolve to the correct ground states. We should also notice that the correct phase boundary and its nature might be concealed by such a biased optimization.

The center panel of Supplementary Fig.\,\ref{fig:initial_dep} presents the magnetizations of the optimized states resulting from the initial FM[111] and FM[100] states, which provide the ground states in the polzarized\,(P) and NP phases, respectively. In the blue\,(red) shaded region, both initial states converge to (almost) identical states such that the magnetizations agree. However, the initial FM[111] state favors the nature of the P state above the phase boundary, whereas the FM[100] state favors the nature of the NP1 state below the phase boundary. Because of these hysteresis-like behavior of the optimization, the phase transition always appears to be first order, since the ground state is identified through energy comparison. Nonetheless, we find that tilting the magnetic field leads to an unambiguous optimization which does not depend on the initial state around the phase boundary between the P and NP phases, as shown in the right panel of Supplementary Fig.\,\ref{fig:initial_dep}. Based on the well converged results, we conclude that the phase transition between the P and NP phase in a tilted field is continuous, which strongly suggests the non-triviality of the NP phases. That is, the NP states may not be smoothly connected to a trivial product state.

\section*{Supplementary Note 3: Bond dimension dependence of Nematic Paramagnet phases }

In this section, we discuss the bond dependence of the nematic paramagnet\,(NP) phases. In most cases, the NP1 phase is obtained by optimizing the initial FM[100] state, where all spins are aligned in the [100]-direction, while the initial FM[011] state leads to the NP2 phase. Supplementary Fig.\,\ref{fig:d_compare} presents the variational energy $E = \langle H \rangle/N$, the size of the magnetization $M = \sqrt{(M^x)^2+(M^y)^2+(M^z)^2}$ and the magnetization components $M^\gamma = (1/N)\sum_i \langle S_i^\gamma\rangle$ for each of the initial FM[100] and FM[011] states with bond dimensions $D=4$, $6$ and $8$. The NP1 phase does not depend strongly on the bond dimension, that is, the spin configurations are already captured well by the $D=4$ ansatz as one can see in the upper right panel of Supplementary Fig.\,\ref{fig:d_compare}. The $D=6$ and $D=8$ ansatze give almost identical energies and magnetizations. On the other hand, the NP2 phase seems to require larger bond dimensions to converge. Notice that the $D=4$ ansatz does not represent the NP2 states well as one can see in the magnetization components\,(right panels of Supplementary Fig.\,\ref{fig:d_compare}). Furthermore, the $D=6$ and $D=8$ states show some discrepancies in the size of magnetization and $z$-component of the magnetization, though their variational energies are quite close each other. We also have found that the variational energies of the zigzag and polarized phases do not change much for $D\geq6$. Therefore, we believe that the phase diagram Fig.\,2 in the main text will be more or less the same should larger bond dimensions are used.

\begin{figure}[!t]
	\includegraphics[width=0.66\textwidth]{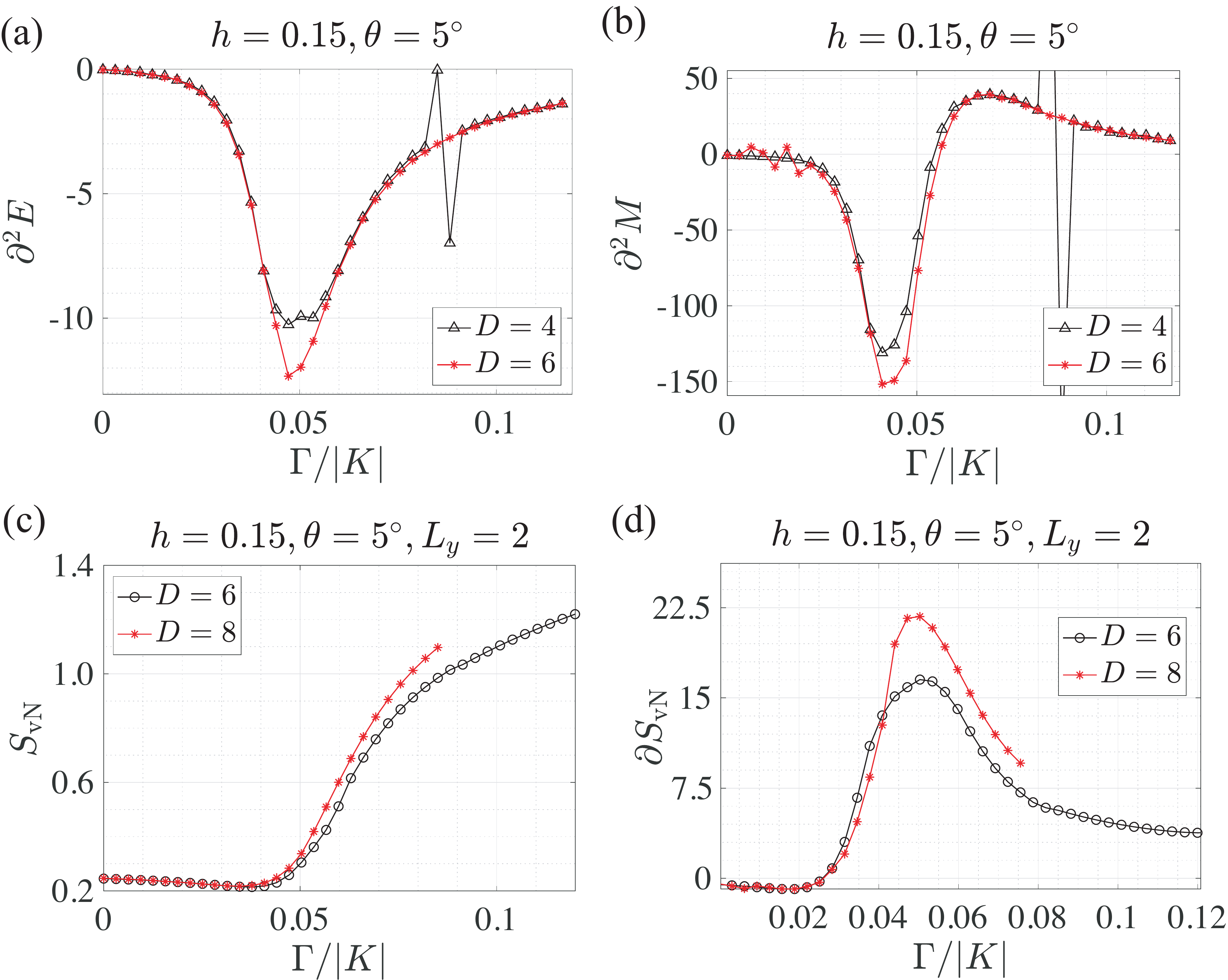}
	\caption{Bond dependence of (a) the second derivative of variational energy, (b) magnetization, (c) the entanglement entropy  and (d) its second derivative at $(\Gamma',h,\theta) = (-0.03,0.15,5^\circ)$ with the tilted field $\theta = 5^\circ$\,(see text). The entanglement entropy is measured on the cylinder geometry with the circumference $L_y=2$.}
	\label{fig:derivative}
\end{figure}

We have also checked the bond dependence of the second derivative of variational energy and magnetization at $(\Gamma', h, \theta) = (-0.03, 0.15, 5^\circ)$, where $\theta$ denotes the tilting angle of the magnetic field\,(see the main text). The results are shown in Supplementary Fig.\,\ref{fig:derivative}\,(a) and (b), and it is clear that the peaks become sharper at larger bond dimension\,($D=6$). The peaks in the first derivative of the entanglement entropy also becomes sharper with larger bond dimension, as shown in Supplementary Fig.\,\ref{fig:derivative}\,(d).

\section*{ Supplementary Note 4: Larger unit-cell magnetic states }
\begin{figure}[!h]
	\includegraphics[width=0.99\textwidth]{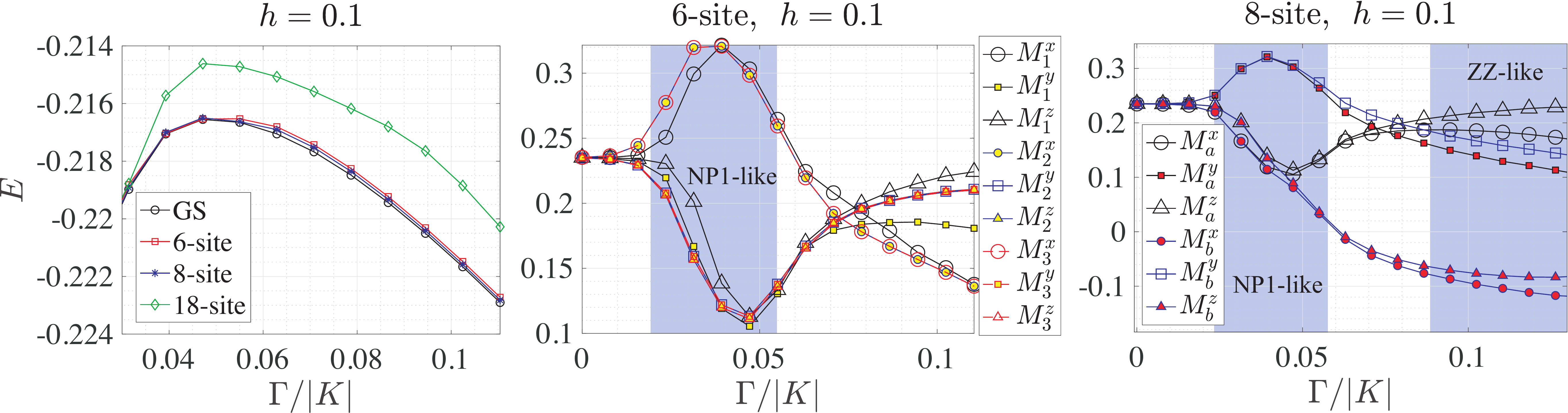}
	\caption{Plots of (left) the variational energies of the large unit-cell anstaze, i.e., the 6-site, 8-site and 18-site1 state, and the magnetization components on (center) three different sublattices of the 6-site order and (right) two different sublattices of the 8-site order.}
	\label{fig:large_unitcell}
\end{figure}

As mentioned in the main text, the larger unit cell magnetic states, e.g., the 6-site, 8-site and 18-site orders found in Supplementary Ref.\,\onlinecite{Li19}, restore the translational symmetry and converge to the polarized state, NP states and even zigzag states after the optimization. On the other hand, they could also be stuck in some local minima, thus their variational energies are significantly higher than the ground state energy. Supplementary Fig.\,\ref{fig:large_unitcell} shows the variational energies obtained from the initial 6-site, 8-site and 18-site magnetic states at $(\Gamma',h) = (-0.03, 0.1)$. As one can see in the left panel, the optimized states from the 18-site order are trapped in bad local minima such that the variational energies are far away from those of the ground states. We have performed the imaginary time evolution\,(ITE) up to $30,000$ steps with $\tau=0.01$\,(see main text). However, the variational energies do not approach the ground state energy throughout the ITE. On the other hand, the 6-site initial states converge to the P and NP states well during ITE by restoring the translational symmetry, i.e., the 2-site unit-cell is recovered. The center panel of Supplementary Fig.\,\ref{fig:large_unitcell} shows that the $M^\gamma_{1,2,3}$, where the subscript denotes the sublattice defined in Supplementary Fig.\,\ref{fig:unitcell}, becomes independent on the sublattice, i.e., the translational symmetry is restored\,(even though it is not perfect). The spin configurations and energy also become similar to the NP1 states obtained from the FM[100] and FM[111] states. Similarly, the 8-site initial state becomes NP1-like states in $0.2 \lesssim \Gamma/|K| \lesssim 0.5 $, while it becomes similar to the zigzag state if $\Gamma/|K| > 0.7$ as shown in the right panel of Supplementary Fig.\,\ref{fig:large_unitcell} such that two zigzag unit cells are realized in the eight sublattices. 

\begin{figure}[!h]
  \includegraphics[width=0.32\textwidth]{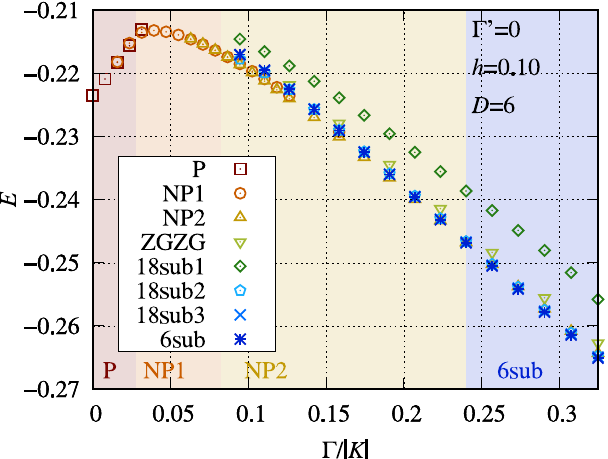}
  \includegraphics[width=0.32\textwidth]{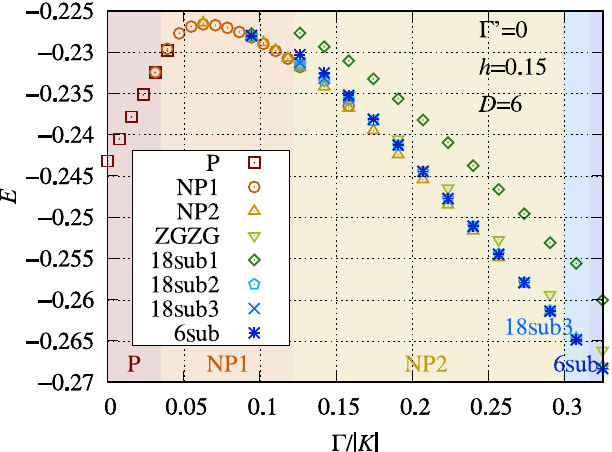}
  \includegraphics[width=0.32\textwidth]{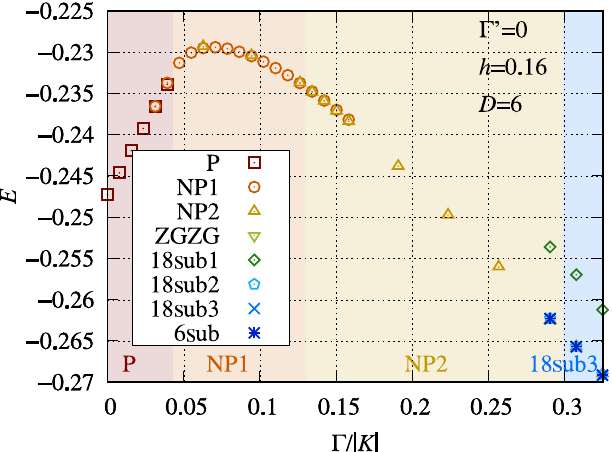}
\\
  \includegraphics[width=0.32\textwidth]{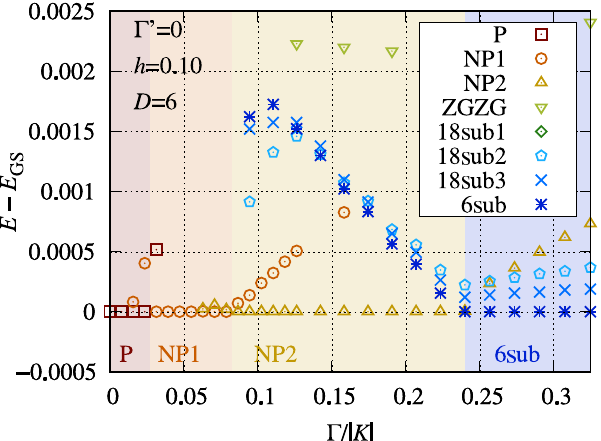}
  \includegraphics[width=0.32\textwidth]{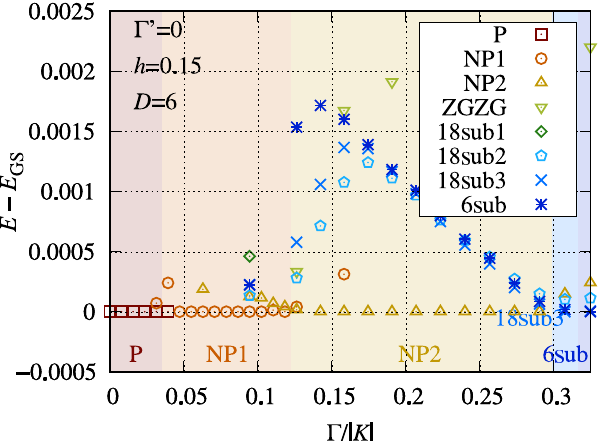}
  \includegraphics[width=0.32\textwidth]{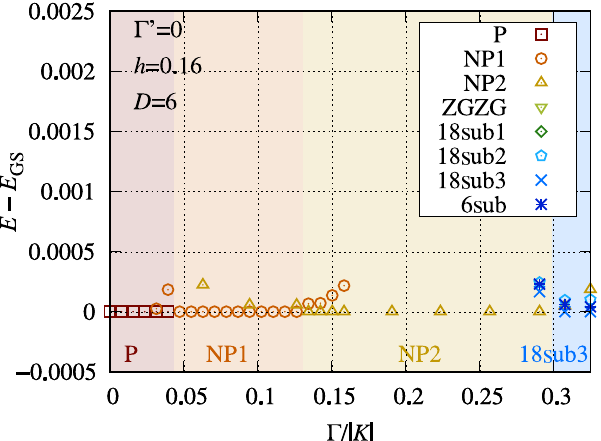}
  \caption{Comparison of (upper) energies and (lower) energy differences
of ground-state candidates for
(left) $h=0.1$,
(center) $h=0.15$, and
(right) $h=0.16$ at $\Gamma'=0$ and $D=6$.}
  \label{fig:large_unitcell_Gp0}
\end{figure}

In contrast, the zigzag state is no longer the ground state at
$\Gamma'=0$, and the larger unit-cell magnetic states appear at
sufficiently large $\Gamma$.
The 6-site and two 18-site orders are competitive,
and the 6-site (18-site) order is favorable at lower (higher) magnetic
fields.
This tendency is consistent with the classical phase
diagram~\cite{Li19}.

\begin{figure}[!h]
	\includegraphics[width=0.99\textwidth]{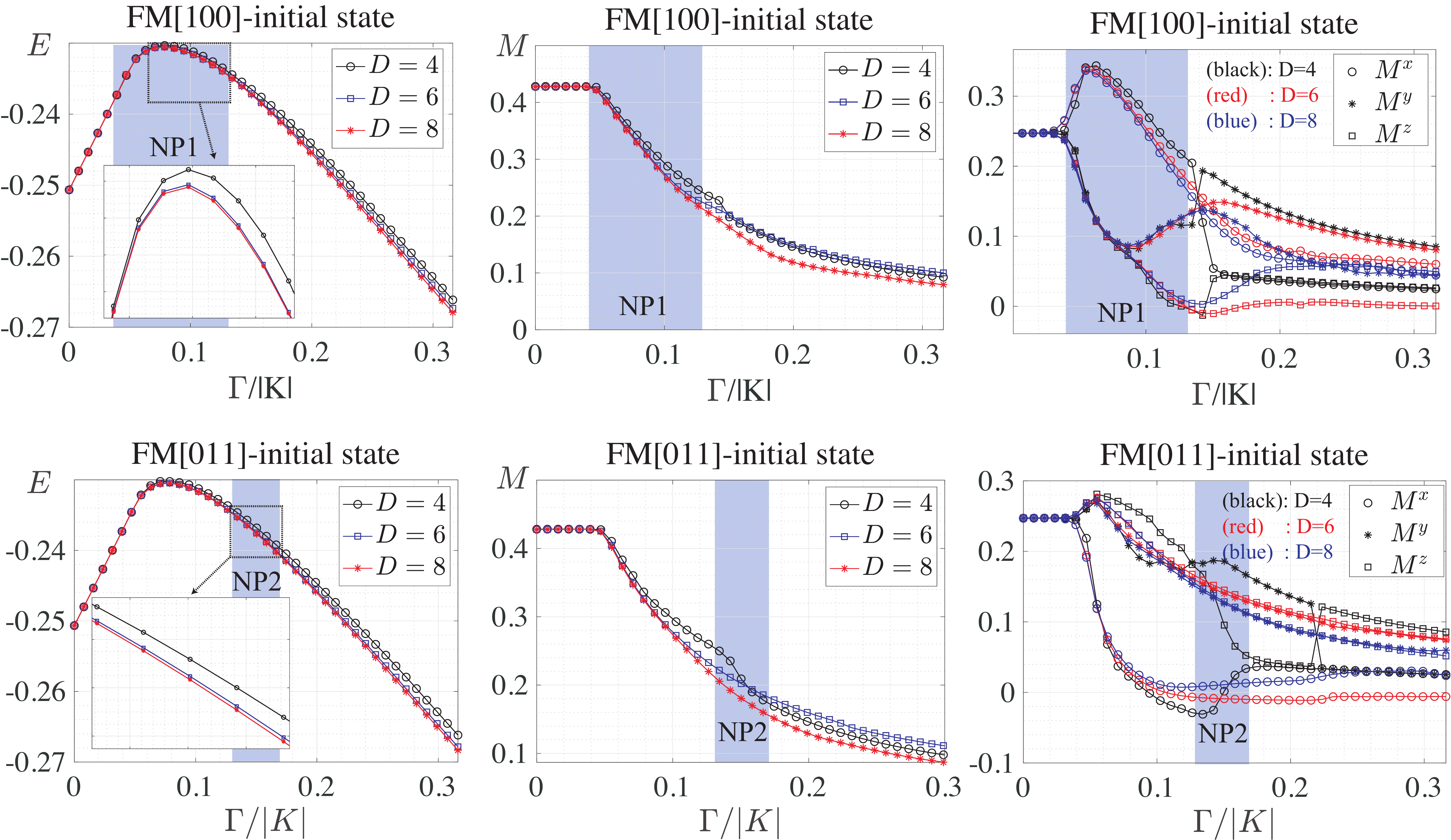}
	\caption{Bond dependence of (left) the variational energy $E = \langle H \rangle/N$, (center) the size of magnetization $M = \sqrt{(M^x)^2+(M^y)^2+(M^z)^2}$ and (right) the magnetization component $M^\gamma = (1/N)\sum_i \langle S_i^\gamma\rangle$ of the optimized states from the initial magnetic (upper) FM[100] and (lower) FM[011] states at $(\Gamma',h) = (-0.03,0.15)$. Here, each optimized state becomes the ground state in the blue shaded region belonging to the NP1 and NP2 phases, respectively.}
	\label{fig:d_compare}
\end{figure}
\section*{ Supplementary Note 5: Critical magnetic field at $\Gamma=0$ }
\begin{figure}[!h]
	\includegraphics[width=0.99\textwidth]{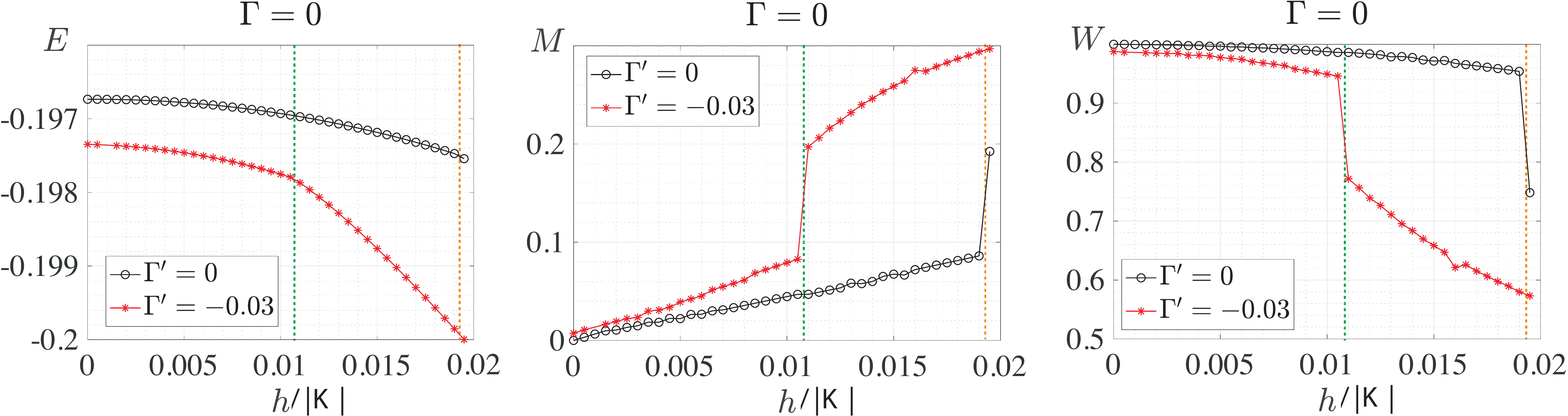}
	\caption{Plots of (left) the variational energy, (center) magnetization and (right) flux expectation value as a function of the field strength $h$ at $\Gamma=0$. The results are obtained from the string gas initial state with bond dimension $D=6$. The yellow and green dotted lines stand for the critical field, at which the phase transition occurs between the Kitaev spin liquid and polarized phases, with $\Gamma'=0$ and $\Gamma'=-0.03$, respectively.}
	\label{fig:hc}
\end{figure}

In this section, we discuss the effect of the $\Gamma'$ interaction on the critical field strength $h_c$, at which a phase transition occurs between the Kitaev spin liquid\,(KSL) and the polarized phase, with $\Gamma=0$. Previous studies reported $h_c \approx 0.02$ from DMRG study\cite{Zhu18} and $h_c \approx 0.025$ from 24-site ED studies\cite{Hickey19, Kaib19} without the $\Gamma'$ interaction. Using the string gas initial state, we have achieved a similar value $h_c \approx 0.01925$ at $\Gamma'=0$, as shown in Supplementary Fig.\,\ref{fig:hc}. The transition seems to be first order at which the magnetization $M$ and the flux expectation value $W$ are discontinuous\,(see the yellow dotted line). The presence of the $\Gamma'$ interaction makes the KSL phase less stable, such that the field induced transition occurs at lower $h$, i.e., $h_c(\Gamma'=-0.03) \approx 0.01075$. See the green dotted line in Supplementary Fig.\,\ref{fig:hc}.

\section*{ Supplementary Note 6: Tilting field toward to the [001]-direction }
\begin{figure}[!h]
	\includegraphics[width=0.85\textwidth]{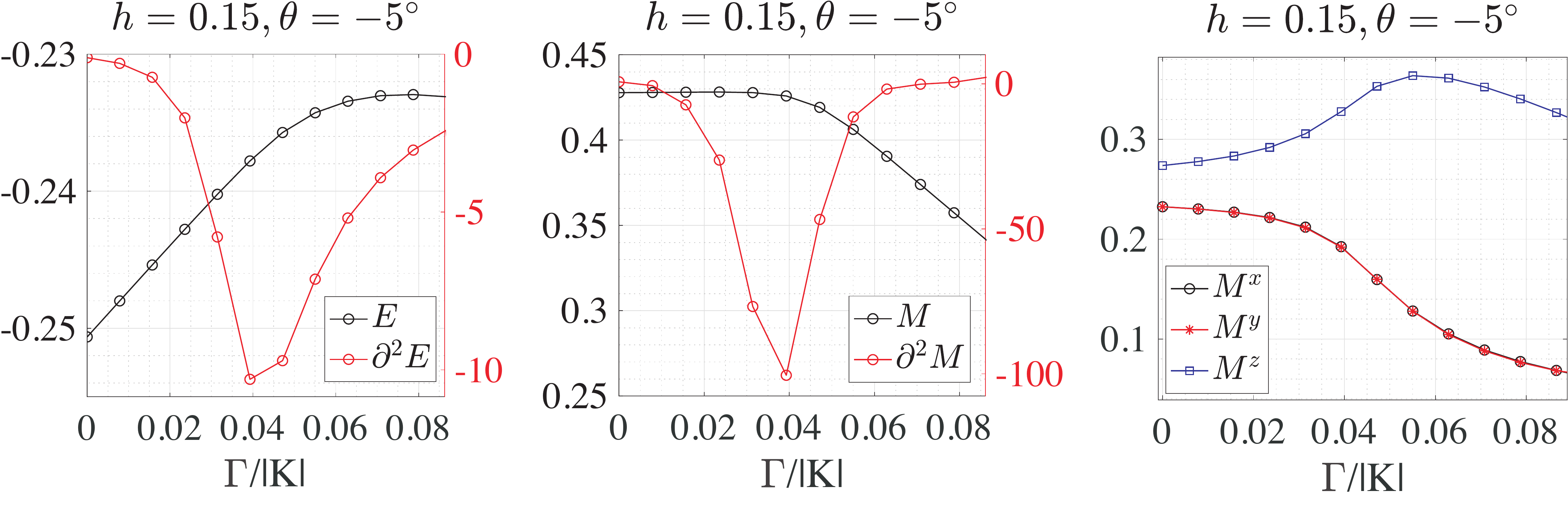}
	\caption{Plots of (left) the variational energy and its second derivative, (center) magnetization and its second derivative, and (right) magnetization components as a function of the field strength $\Gamma/|K|$ at $(\Gamma',h,\theta) = (-0.03, 0.15,-5^\circ)$. The results are obtained from the $D=6$ ansatze.}
	\label{fig:opposite_tilt}
\end{figure}

In the main text, we consider tilting the field towards the $[11\bar{2}]$-direction\,($\theta$: tilting angle), which leads to the phase transition between the polarized and NP2 phases without going through the NP1 phase, as shown in Fig.\,4. Here, we present the results of tilting the field towards the [001]-direction, i.e., $\theta = -5^\circ$, in Supplementary Fig.\,\ref{fig:opposite_tilt}. It leads to the continuous-like phase transition between the polarized and NP1 phases as one expect.

\section*{ Supplementary Note 7: Magnetization of the nematic paramagnet phase }
\begin{figure}[!h]
	\includegraphics[width=0.6\textwidth]{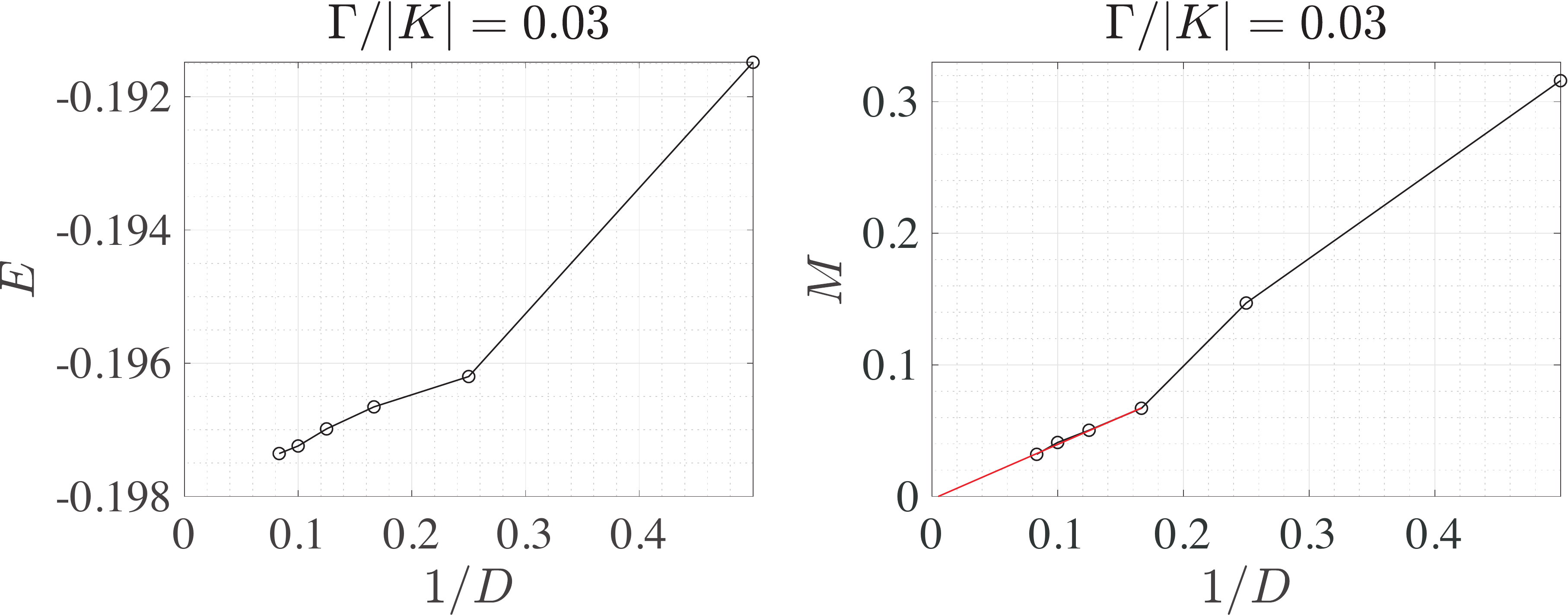}
	\caption{Plots of (left) the variational energy and (right) magnetization as a function of the inverse of the bond dimension $D$ at $\Gamma/|K| = 0.03$ of the $K$-$\Gamma$ model, i.e., $(\Gamma',h) = (0,0)$. The red solid line is the linear fitting curve for the data with $D=6,8,10$ and $12$.}
	\label{fig:kg_scaling}
\end{figure}

In the $K$-$\Gamma$-$\Gamma'$ model, the NP phases appear at finite magnetic fields and are not stabilized in the zero field limit. Without the $\Gamma'$ interaction, the NP phases become wider and survive down to almost zero field. Here, we show that the NP2 state becomes non-magnetic at zero field. Even though the NP states seem to have spontaneous magnetizations, it is an artifact of the finite bond dimension $D$. Supplementary Fig.\,\ref{fig:kg_scaling} presents the $D$-scaling of the variational energy and magnetization at $\Gamma/|K| = 0.03$ of the $K$-$\Gamma$ model. As one can see, even with $D \geq 10$ which is a considerably large bond dimension, the energy and the magnetization exhibit a visible evolution instead of a convergent behavior. To extrapolate the magnetization at $D \rightarrow \infty$, we have fitted the magnetization at $D=6,8,10$ and $12$ with a linear function (see the red solid line in the center panel). It strongly indicates a zero magnetization at $D\rightarrow\infty$, and thus the NP phases only develop finite magnetizations in the presence of an external field.

\section*{ Supplementary Note 8: Energy comparison between FM and NP phases }
\begin{figure}[!h]
	\includegraphics[width=0.99\textwidth]{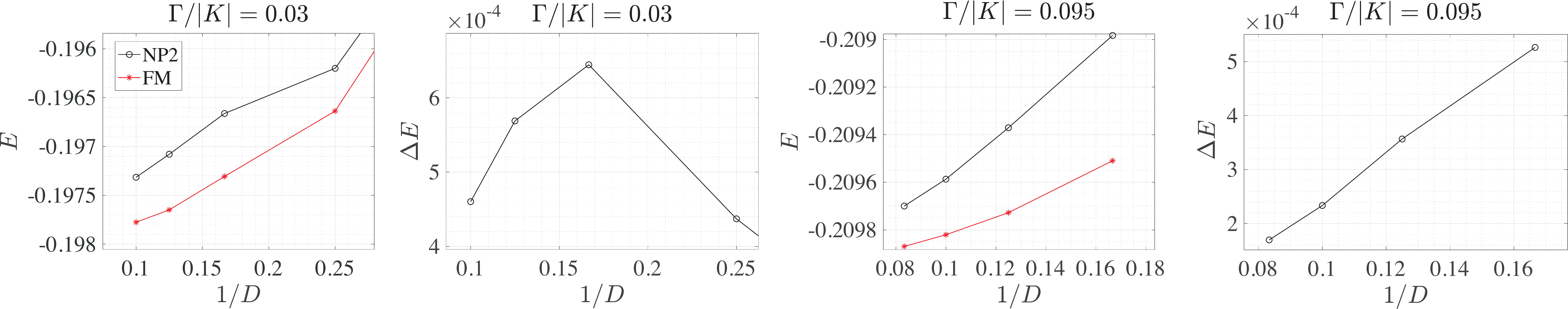}
	\caption{Plots of variational energies $E$ and their differences $\Delta E = E_{\rm NP2} - E_{\rm FM}$ of the NP and FM states at $\Gamma/|K| = 0.03$ and $0.095$.}
	\label{fig:fm_np_compare}
\end{figure}

As shown and discussed in the main text, both $K$-$\Gamma$ and $K$-$\Gamma$-$\Gamma'$ model exhibit a ferromagnetic phase, where all spins align in the $[1\bar{1}\bar{1}]$-direction, in the zero field limit\,(see Figs.\,1 and 5 in the main text). However, the variational energies between the FM and NP2 states are very close, i.e., the energy difference $\Delta E = E_{\rm NP2} - E_{\rm FM} \sim O(10^{-4})$, as shown in the right panel of Supplementary Fig.\,\ref{fig:fm_np_compare}. The results strongly suggest the possibility that the NP2 state may be preferred over the FM states at $h=0$. This might be captured with much larger bond dimensions.

\section*{ Supplementary Note 9: Phase transition between the chiral Kitaev spin liquid and the nematic phase in the tilted magnetic field }
\begin{figure}[!h]
	\includegraphics[width=0.99\textwidth]{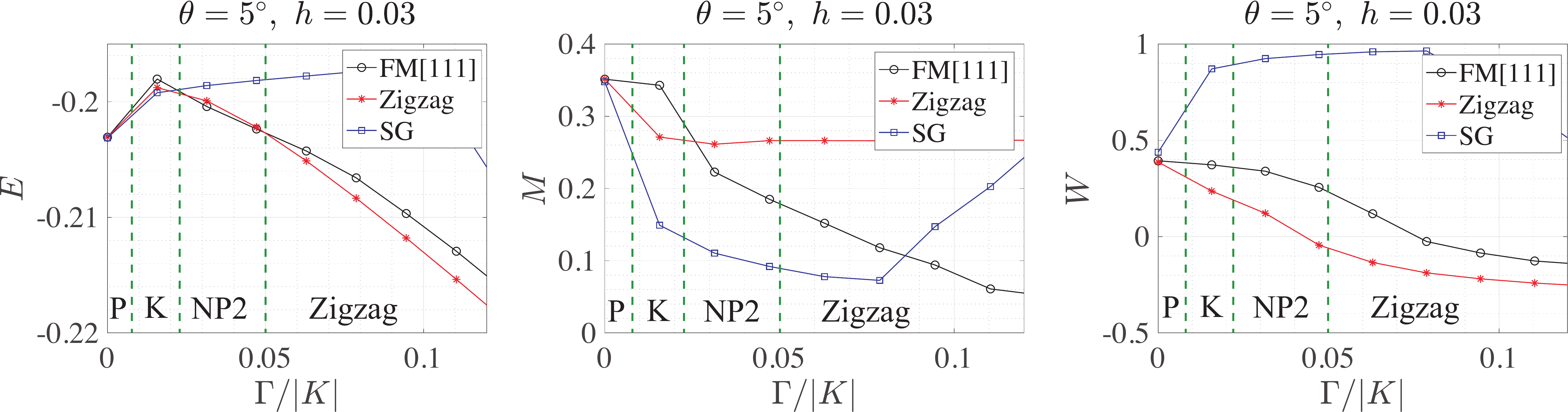}
	\caption{Plots of (left) the variational energy, (center) the magnetization, and (right) the flux expectation value of the optimized states from the initial FM[111], Zigzag and SG states. The green dotted lines stand for the phase boundaries determined by comparing the variational energies. Here, K denotes the chiral KSL phase.}
	\label{fig:tilt_phase}
\end{figure}

Without tilting the field\,$(\theta = 0)$, the KSL and the NP phases are clearly distinct each other as shown and discussed in the main text. Since the NP phases are also non-magnetic, one may wonder if those are qualitatively different even with the tilted field where the rotational symmetry is explicitly broken. We have found that those two phases are still distinct each other with the tilted field\,($\theta=5^\circ$). Figure \ref{fig:tilt_phase} presents the variational energy, the magnetization and the flux expectation value of the optimized states from the initial FM[111], zigzag and SG states at $h=0.02$ and $\theta = 5^\circ$. The initial FM[111] state flows into the NP2 states well, while the initial SG state keeps the KSL nature in $0.01 \lessapprox \Gamma/|K| \lessapprox 0.07 $. As in the case of $\theta=0$, the results clearly show that there are successive phase transitions between the polarized and zigzag phases: P $\rightarrow$ KSL $\rightarrow$ NP2 $\rightarrow$ Zigzag. In particular, the KSL and NP2 states give certainly distinct magnetizations and flux curves in each phase.

\section*{ Supplementary Note 10: Coefficients of the area law in the polarized and NP2 phases. }

The entanglement entropy in the P and NP2 phases at $h=0.15$ and $\theta = 5^\circ$ strictly follows the area law, i.e., $S_{\rm vN} \simeq \alpha L_y$. Here, we show the coefficient $\alpha$ directly in Supplementary Fig.\,\ref{fig:ee_L}. It is almost constant in the P phase while gradually increases with $\Gamma$ in the NP2 phase.

\begin{figure}[!h]
	\includegraphics[width=0.3\textwidth]{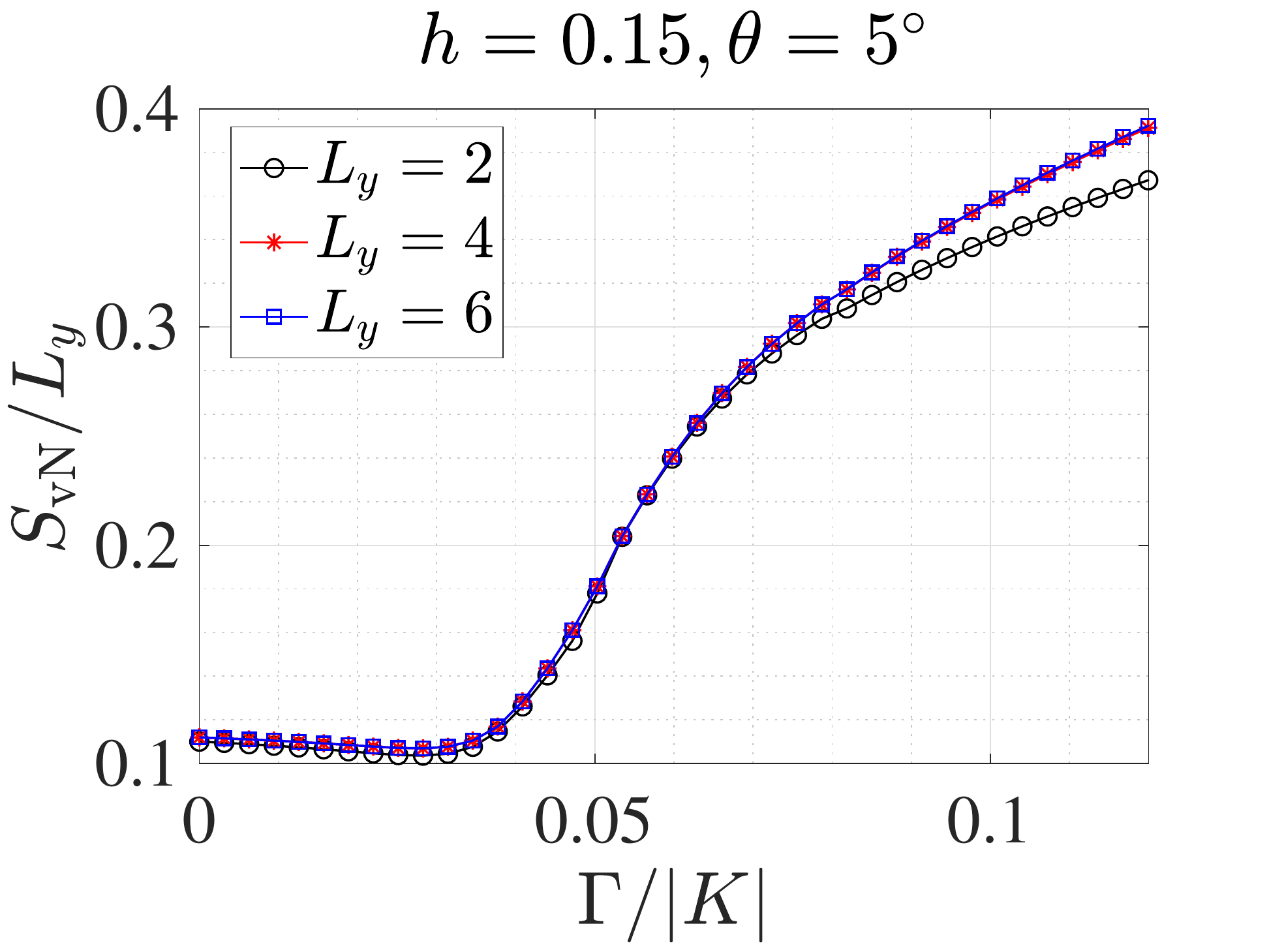}
	\caption{The coefficient of the area law in the entanglement entropy at $h=0.15$ and $\theta = 0.15^\circ$.}
	\label{fig:ee_L}
\end{figure}
%


%

\end{document}